# Collective spin excitations in bi-component magnonic crystals consisting of bi-layer Permalloy/Fe nanowires.

G. Gubbiotti,[1] S. Tacchi,[1] M. Madami,[2] G. Carlotti,[2,3] Z. Yang,[4] J. Ding,[4] A. O. Adeyeye,[4] and M. Kostylev[5]

[1]Istituto Officina dei Materiali del Consiglio Nazionale delle Ricerche (IOM-CNR), Sede di Perugia, c/o Dipartimento di Fisica e Geologia, Via A. Pascoli, I-06123 Perugia, Italy.

[2]Dipartimento di Fisica e Geologia, Università di Perugia, Via A. Pascoli, I-06123 Perugia, Italy.

[3]Centro S3, c/o Istituto Nanoscienze del CNR (CNR-NANO), I-41125 Modena, Italy

[4]Information Storage Materials Laboratory, Department of Electrical and Computer Engineering, National University of Singapore, 117576 Singapore.

[5]School of Physics, University of Western Australia, Crawley 6009, WA, Australia.

In the developing field of magnonics, it is very important to achieve tailoring of spin wave propagation by both a proper combination of materials with different magnetic properties and their nanostructuring on the submicrometric scale. With this in mind, we have exploited deep ultra-violet lithography, in combination with tilted shadow deposition technique, to fabricate arrays of closely spaced bi-layer nanowires (NWs), with separation d=100 nm and periodicity $a$=440 nm, having bottom and top layers made of Permalloy and Iron, respectively. The NWs have either "rectangular" cross section (bottom and upper layers of equal width) or "L-shaped" cross section (upper layer of half width). The frequency dispersion of collective spin wave excitations in the above bi-layered NWs arrays has been measured by the Brillouin light scattering technique while sweeping the wave vector perpendicularly to the wire length over three Brillouin zones of the reciprocal space. For the rectangular NWs, the lowest-frequency fundamental mode, characterized by a quasi-uniform profile of the amplitude of the dynamic magnetization across the NW width, exhibits a sizeable and periodic frequency dispersion. A similar dispersive mode is also present in L-shaped NWs, but the mode amplitude is concentrated in the thin side of the NWs. The width and the center frequency of the magnonic band associated with the above fundamental modes has been analysed, showing that both can be tuned by varying the external applied field. Moreover, for the L-shaped NWs it is shown that there is also a second dispersive mode, at higher frequency, characterized by an amplitude concentrated in the thick side of the NW. These experimental results have been quantitatively reproduced by an original numerical model that includes two-dimensional Green's function of the dipole field of the dynamic magnetization and interlayer exchange coupling between the layers.

**Introduction**

Bi-component magnonic crystals (BMCs) consisting of two different periodically arranged magnetic materials have attracted great interest in recent years because of the possibility of tuning the spin wave (SW) band structure, such as the frequency position and the width of the allowed band and forbidden bandgap,[1,2,3,4,5,6,7] thanks to the additional degrees of freedom offered by the contrasting properties of the ferromagnetic (FM) materials.

In one-dimension (1D), BMCs have been experimentally demonstrated, in the form of periodic arrays of alternating contacting Cobalt (Co) and Permalloy (Py) nanowires (NWs). The SW dispersion relations were mapped for both the parallel[8,9] and antiparallel[10] phases and the dependence of the bandgap width and frequency position on the external magnetic field have been investigated. These studies suggest that there is a strong direct exchange coupling across the Co-Py interface which influences the pinning of the dynamical magnetization at the interface.[10,11]

In two-dimensions (2D), NiFe antidot lattices with embedded Co dots[12,13,14] and array of alternated Py and Co nanodots have been investigated.[15] The studies on 2D BMCs reveal that their band structure is very rich due to the large density of modes and their consequent hybridization. Here, additional complexity with respect to the 1D case derives from the pronounced non-uniformity of the internal static field due to static demagnetization effects.[16]

To fully exploit BMCs' capabilities in applications, however, the key challenge is associated with their nanofabrication. There are various limitations to the quality of BMCs produced with the multilevel electron beam lithographic approach including the alignment between successive fabrication steps which can create thin oxidized layers at the interfaces between ferromagnetic materials.[10]

In this work, the SW dispersion (frequency versus wave vector) in BMCs consisting of arrays of bi-layered (Py-Fe) NWs has been measured by the Brillouin light scattering (BLS) technique thanks to its wave vector sensitivity. All the samples have fixed width ($w_2$) of the Py NWs and differ by the width of the Fe ($w_1$) ones. For NWs with "rectangular" cross section the bottom Py and upper Fe layers are of equal width $w_1 = w_2 = 340$ nm, while for "L-shaped" NWs $w_1 = 170$ nm and $w_2 = 340$ nm. These samples have been fabricated by the self-aligned shadow deposition technique[17] which, unlike the multilevel electron beam lithographic process, does not require the alignment between the two FM layers. In addition, the deposition of the two materials can be performed without breaking the vacuum in the same process step thus ensuring high quality of the interface between the two FM materials. This set of NWs has been previously studied with the ferromagnetic resonance (FMR) method in order to understand the impact of the strongly broken symmetry of such nanostructures on the FMR response.[18] However, with the above technique, it was only possible to take measurements for the Γ-point of the 1st Brillouin zones (BZ), i.e. at a wave vector $q=0$, where magnetization dynamics represents a family of in-plane standing spin waves with zero Bloch wave numbers. Here instead, the BLS analysis was performed by sweeping the wave vector perpendicularly to the wire length over three BZs of the reciprocal space. Remarkably, for all the investigated NWs the fundamental mode, lying at the lowest frequency, shows the largest frequency dispersion. The magnetic field dependence of both the width and the center of the corresponding magnonic band has been investigated, showing that both depend on the number of layers and on the width of the overlying Fe NWs. Moreover, for the L-shaped cross-section NWs, it is shown that there is also another, high-order, mode with a sizeable dispersion. Quantitatively, the experimental results have been successfully reproduced by numerical simulations based on an original model which includes two-dimensional Green's function description of the dynamic dipole field of the dynamic magnetization and the interlayer exchange coupling between the layers. This approach also enabled us to calculate the spatial profile of the dynamical magnetization corresponding to each of the detected eigen-modes, showing that those that exhibit a sizeable dispersion are characterized by the absence of nodes across the NW width.

**Sample fabrication and Brillouin light scattering measurements**

Large area (4x4 mm$^2$) arrays of bi-layer NWs arrays were fabricated on silicon substrates by using deep ultra-violet lithography in combination with recently developed tilted shadow deposition technique.[17] Five NWs arrays, whose geometrical and magnetic parameters are summarized in Table I, have been studied. In all investigated NWs arrays, the lower NWs is formed by Permalloy (Ni$_{80}$Fe$_{20}$, Py); it has a fixed width $w_2$ =340 nm and a thickness $L_2$ =10 nm. The Py NWs were capped with Fe NWs having a width $w_1$ = 340 nm and a thickness $L_1$ =10 nm ("Rectangular" cross-section) or $w_1$ = 170 nm ("L-shaped" cross-section) and a thickness $L_1$ =10 or 20 nm. Single layer NWs of either Py ($w_2$=340 nm, $L_2$=10 nm) or Fe ($w_1$=340 nm, $L_1$=10 nm) were fabricated and used as reference samples. For all the arrays, the inter-wire edge-to edge distance was $d$=100 nm (as measured at the level of Py layers) and the array periodicity $a$= ($w_2$+$d$) = 440 nm, resulting in the edge of the first BZ located at $\pi/a$ = 0.71 ×10$^7$ rad×m$^{-1}$. The geometric details of these samples are listed in the third and fourth columns of Table 1.

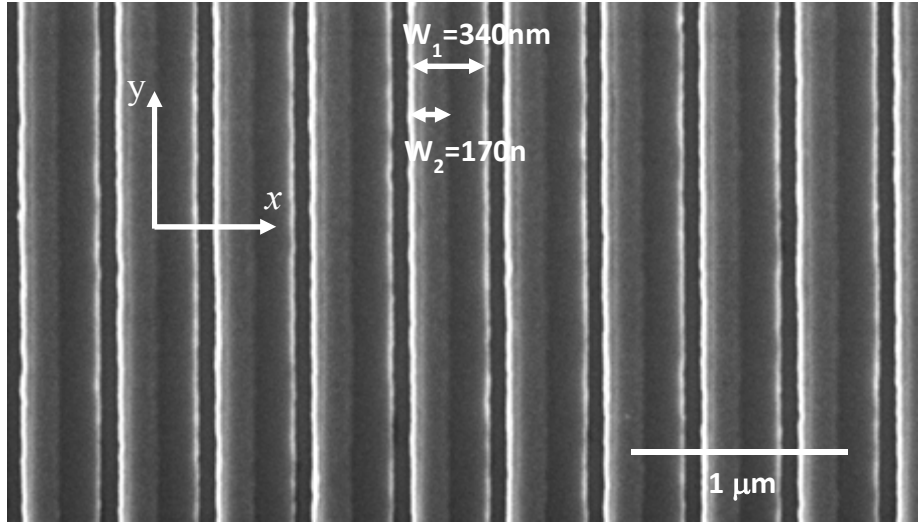

**Fig. 1** SEM micrographs of the L-shaped NW array (Sample # 4) together with the coordinate system used in the calculation. NW magnetization is aligned along the *y*-direction.

BLS spectra were recorded at room temperature in the back-scattering configuration by applying a magnetic field of H=500 Oe along the NWs length (*y*- direction) and sweeping the wave vector q in the orthogonal direction (*x*- direction).[19] At this field the NWs are saturated along the *y*-direction, as inferred from the measured magneto-optic Kerr effect hysteresis loops (not shown here). About 200 mW of *p*-polarized monochromatic light from a solid state laser λ=532 nm was focused onto the sample surface for different incidence angles of light with respect to the sample normal and the *s*-polarized backscattered light was analyzed with a (3+3) tandem Fabry-Pérot interferometer.[20] The magnitude of the in-plane wave vector is linked to the incidence angle of light $\theta_i$ and to the light wavelength λ by the following relation: $q = (4\pi/\lambda) \times \sin\theta_i$.

Table I. Geometric and magnetic parameters for the single- and bi-layered NWs.

| Sample label | Number of layers and used materials | Layer Width $w_1$-$w_2$ (nm) | Layer Thickness $L_1$-$L_2$ (nm) | Gyromagnetic ratio (MHz/Oe) | Saturation magnetization $4\pi M$ (G) | Exchange constant A ($10^{-6}$ erg/cm) | Interlayer Exchange constant $A_{12}$ (erg/cm$^2$) |
|---|---|---|---|---|---|---|---|
| #1 | 1, Py | 340 | 10 | 2.9 | 7000 | 1 | N/A |
| #2 | 1, Fe | 340 | 10 | 3.05 | 12500 | 2 | N/A |
| #3 | 2, Fe-Py "rectangular" | 340-340 | 10-10 | 3.05-2.9 | 16500-10000 | 2-1 | 6 |
| #4 | 2, Fe-Py "L-shaped" | 170-340 | 10-10 | 3.05-2.9 | 18500-9000 | 2-1 | 6 |
| #5 | 2, Fe-Py "L-shaped" | 170-340 | 20-10 | 3.05-2.9 | 18000-7500 | 2-1.2 | 6 |

**Theory**

In order to interpret the dispersion of collective SW excitations and the cross-sectional profiles of the modes, we used a numerical method which is a further development of the quasi-analytical approach from Kostylev et al.[21] It is based on solution of the linearized Landau-Lifshitz-Gilbert (LLG) equation in the magnetostatic approximation with 2D Green's function description of the dipole field of dynamic magnetization.[18] As for any periodic medium, the spin waves on periodic arrays of interacting NWs represent Bloch waves. This implies that the distribution of their vector amplitude across the crystal unit cell $a$ is given by a spatially periodic function ("Bloch function") $\tilde{\mathbf{m}}(x,z,q)$, ($\tilde{\mathbf{m}}(x,z,q) = \tilde{\mathbf{m}}(x+a,z,q)$), and the variation of the phase of this distribution from unit cell to unit cell by a Bloch wave number $q$.[22]

$$\mathbf{m}(x,z) = \tilde{\mathbf{m}}(x,z,q)\exp(iqx), \quad (1)$$

Accordingly, the dispersion $\omega(q)$ of those waves is periodic with the first Brillouin zone spanning from $q=-\pi/a$ to $q=+\pi/a$.

As we consider NWs magnetized along their longitudinal axes, the static magnetization configuration for the material is uniform, with all spins in both layers pointing along this direction. To calculate the dynamic dipole field $\mathbf{h}_{dip}(x,z)$ created by the dynamic magnetization we use the Green's function approach:

$$\mathbf{h}_{dip}(x,z) = \int_V \hat{G}(x-x',z-z')\mathbf{m}(x,z)dV, \quad (2)$$

where $\hat{G}(s,p)$ is given by Eqs.(1-2) in Ref. [22] and $V$ is the volume of the magnetic material. The presence of the Green's function in the expression for the total dynamic effective field $\mathbf{h}$ results in an integral part for the expression. Since the sizes of the NW cross-section are comparable with the exchange length of the magnetic materials, the contribution from the effective exchange field of dynamic magnetization is also included. Based on previous studies,[23,24] "unpinned surface spins" boundary conditions[25] are assumed at all NWs surfaces. The exchange field is given by the usual differential operator (Eq.(2) in Ref.[34])

$$\mathbf{h}_{exc}(x,z) = \alpha_i(\partial^2/x^2 + \partial^2/z^2)\mathbf{m}(x,z), \quad (3)$$

where $\alpha_i = A_i/(2M_i^2)$ is the exchange constant for the $i$-th layer ($i=1,2$) having saturation magnetization $M_i$.

Both the integral and the differential parts of the operator for **h** are discretized on a square mesh which fills the NW cross-section. This procedure is the same as previously employed in Refs. 4 and 5. In this work an additional contribution is considered because of the presence of two exchange-coupled ferromagnetic materials. They possess different values of saturation magnetization ($M_1$ and $M_2$), intra-layer exchange constant ($A_1$ and $A_2$), and gyromagnetic ratio ($\gamma_1$ and $\gamma_2$), above and below the layer interface, respectively. On top of this, the interface exchange boundary conditions for the dynamic magnetization are included in the model. We use the linearized boundary conditions:[26]

$$\partial m_x^{(1)} / \partial z + \frac{A_{12}}{A_1} m_x^{(1)} - \frac{A_{12}}{A_1} \frac{M_1}{M_2} m_x^{(2)} = 0$$
$$\partial m_x^{(2)} / \partial z - \frac{A_{12}}{A_2} m_x^{(2)} + \frac{A_{12}}{A_2} \frac{M_2}{M_1} m_x^{(1)} = 0$$
(4)

In this expression, $m_x^{(1)}$ and $m_x^{(2)}$ are the in-plane ($x$) components of the dynamic magnetization in Layers 1 (Fe) and 2 (Py) respectively while $A_{12}$ is the inter-layer exchange constant. A similar boundary condition is used for the perpendicular-to-plane (i.e. perpendicular-to-the-interface) component of dynamic magnetization $m_z$. As in Ref. 27, the surface and interface boundary conditions are included into the boundary elements of the discrete (finite-difference) version of the exchange operator. In our previous work on Py L-shaped NWs array,[21] the interlayer exchange coupling was not considered because of the presence of one FM material, only.

The discretization of the operators taking into account (1) transforms the lineariized LLG equation

$$i\omega \mathbf{m} = -\gamma_i \left[ \mathbf{m} \times \mathbf{H} + M_i \mathbf{h} \times \mathbf{e}_y \right]$$ (5)

into an eigenvalue problem for a matrix:

$$i\omega | \tilde{m} >= \hat{C} | \tilde{m} >.$$ (6)

Here $\omega = 2\pi f$ is circular frequency, $\gamma_i$ is the gyromagnetic ratio for the i-th layer, $\mathbf{e}_y$ is the unit vector in the direction of the static magnetization (y-direction), $| \tilde{m} >$ is a block column-vector with blocks $(\tilde{m}_{xj}, \tilde{m}_{zj})$, where $\tilde{m}_{\lambda j}$ is the λ-component (i.e. $x$ or $z$ component) of the Bloch function for the point $j$ of the mesh, and $\hat{C}$ is the matrix obtained by discretizing the LLG equation after substitution of (1) into it. In Eq. (3) the Bloch wave eigen-frequencies correspond to the eigen-values of $\hat{C}$ while $\tilde{\mathbf{m}}(x,z,q)$ are the cross-sectional profiles of the modes corresponding to the different eigen-frequencies.

Solutions of Eq. (6) are obtained by using numerical tools built in into commercial MathCAD software. For the mesh size 3.33×3.33 nm$^2$, it takes about 5 minutes to obtain all eigen-values of $\hat{C}$ with a quad-core personal computer for the most computation-demanding geometry – that of the L-shaped NW with the 20 nm-thick Fe layer. The determination of the magnetic parameters through the fitting procedure has been considered as acceptable when the frequency difference between the calculated and measured frequency of the whole set of modes is smaller than ±0.5 GHz, the maximum uncertainty in the determination of the spin wave frequency in the measured BLS spectra. The values of the magnetic parameters which provide the best agreement of the simulation results with the experimental data are listed in the last four columns of Table I. It appeared that keeping the values for the gyromagnetic ratio and the intra-layer exchange constant (except for the Py layer of Sample #5) equal to the literature values[18] and varying the remainder of the parameters – the saturation magnetization and the inter-layer exchange constant, allows obtaining reasonable agreement with the experiment.

Similar to our FMR study of the same materials,[18] we found that the calculated mode frequencies are highly sensitive to assumed saturation magnetization values for the layers. For instance, a decrease in $4\pi M$ for Py from 8500 G to 7500 G shifts the fundamental mode for Sample #5 as a whole downwards by 400 MHz. Simultaneously, the frequency of the third mode is shifted down by 750 MHz.

The saturation magnetization values for the materials as extracted from the best fits (Table 1) are significantly smaller than the generally accepted ones for those materials. For instance, for the Py layer of Sample #5 we obtained 7500 G instead of 10000 G. This is because the extracted values represent the "effective" saturation magnetization that includes, for instance, the effect of the surface perpendicular magnetic uniaxial anisotropy (PMA)[28] that may be originated by surface oxidation.[29] Its presence is corroborated by the fact that the Py layer of Sample #3, the only one completely protected from oxidation by the presence of the Fe overlayer, the obtained value of the saturation magnetization is similar to the expected value for bulk Py and larger than in L-shaped samples.

To go deeper into the effect of the presence of PMA, one should notice that the amount of frequency downshift of Damon-Eshbach spin waves caused by PMA depends on the wave number. Indeed, from Fig. 4 in the work of Kalinikos et al.,[30] one finds that as easy-axis anisotropy decreases the slope of the dispersion curve. On the other hand, the wave number dependence of the PMA induced frequency shift for a Damon-Eshbach wave in a continuous film implies that, in the present case of a confined geometry, PMA will affect differently different collective modes and different $q$-values. Including the effective field of PMA in our numerical model is a trivial task; we made a number of trial simulations for which we kept the value of the saturation magnetization for Py equal to a standard value $4\pi M_1$=10 kG and included a non-vanishing effective field of PMA $H_{PMA}$. We found that, although it is possible to obtain a good fit for the frequency for the fundamental mode for $H_{PMA}$ =2600 Oe, the second *dispersive* mode and its nearest neighbor from below moved prohibitively low in frequency with respect to the experimental data of Fig. 4(b). This agrees with the above-mentioned fact that the presence of PMA leads to a smaller dispersion slope for the Damon-Eshbach wave. However, this might also suggest that PMA is not the only contributor to the small values of the effective saturation magnetization; additional contribution can in principle be due to bulk in-plane anisotropy of the layers. Also, one may expect that for the L-shaped NWs, the anisotropy at the Py/iron interface may differ from the surface anisotropy of the part of the Py layer not covered by iron (see the discussion of the oxidation effect above). It would not be difficult to include all these anisotropies into the numerical model; however, because of the significant time for a program run, their inclusion makes the material parameter space too large for performing a numerical best-fit procedure. Given all these considerations, we use the simple approach of the effective saturation magnetizations (Table 1) with the aim of providing a physical explanation for the complicated BLS spectra observed for this novel nanostructure, while a detailed investigation of magnetic parameters dependence on the sample structure is beyond the scope of this work.

**Results and Discussion**

  *a) Dispersion relation and spatial profile of the fundamental mode for NWs with rectangular cross-section*

In Fig. 2, we present a sequence of representative BLS spectra for Sample #3, i.e. Py/Fe bi-layer NWs with rectangular cross section, measured at $q=n\pi/a$ *(with n=0,1,2 and 3)*. Spectra have a very good signal-to-noise ratio and are characterized by the presence of several well-resolved peaks, with the most intense peak which, on increasing the wave vector, gradually moves toward higher order modes, as shown in previous investigations of magnetic wires[31] and dots.[32] It can be seen that the lowest-frequency fundamental mode, whose intensity is maximum at $q=0$, exhibits a remarkable frequency evolution with $q$, while the modes at higher frequency are much less dispersive, in agreement with previous investigations of SW in arrays of interacting stripes.[33]

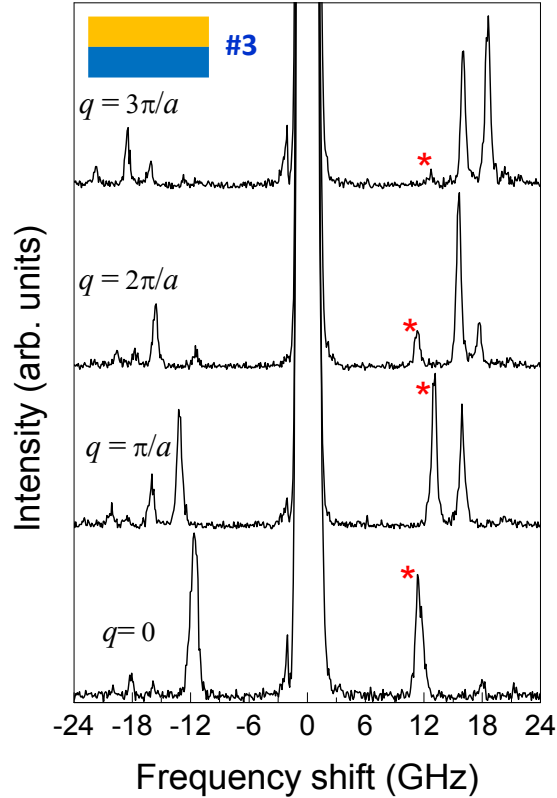

**Fig. 2** (Color online) BLS spectra for Sample #3 measured for H=500 Oe applied along NW length (easy direction) at different wave vector values $q=n\pi/a$ (with $n$=0,1,2 and 3). The lowest frequency mode, characterized by a significant frequency variation, is marked by the red asterisk on the anti-Stokes side of the spectra. Insets show the schematic cross-sectional view of the NWs (blue is Py and orange is Fe).

In Fig. 3 (a) we compare the frequency dispersion of the lowest frequency dispersive mode for Samples #1-3. The dispersion is periodic with the appearance of BZs determined by the artificial periodicity of the stripes array. The periodicity of the frequency oscillation (width of the Brillouin zone) is independent of the thickness of the layers, since all the investigated NWs arrays have the same lattice period $a$. Concerning the center position of the magnonic band, one sees that it is up–shifted passing from Sample #1 to #3, i.e. from the single layer Py NWs to the single Fe layer and finally to the bi-layered NWs. In addition, also the width of the magnonic band (frequency variation between $q$= 0 and $\pi/a$) is more pronounced for bi-layered NWs than for single layer ones. In all the cases, the experimental data are well reproduced by theoretical calculations, performed by using the magnetic parameters reported in Table I. Moreover, from the calculations one can obtain the spatial profiles of the considered mode for the different samples, as shown in Fig. 3(b). The profiles are calculated at either $q=0$ or $q=\pi/a$ and the dynamic magnetization distributions have been averaged over the thicknesses of the respective layers to create the two-dimensional plots. It can be seen that the spatial profile is typical of a fundamental mode, without any node across the NWs width and maximum amplitude in the center. This fundamental mode creates a large dynamic dipolar field which efficiently couples a NW with its neighbors, resulting in a sizeable dispersion.[34]

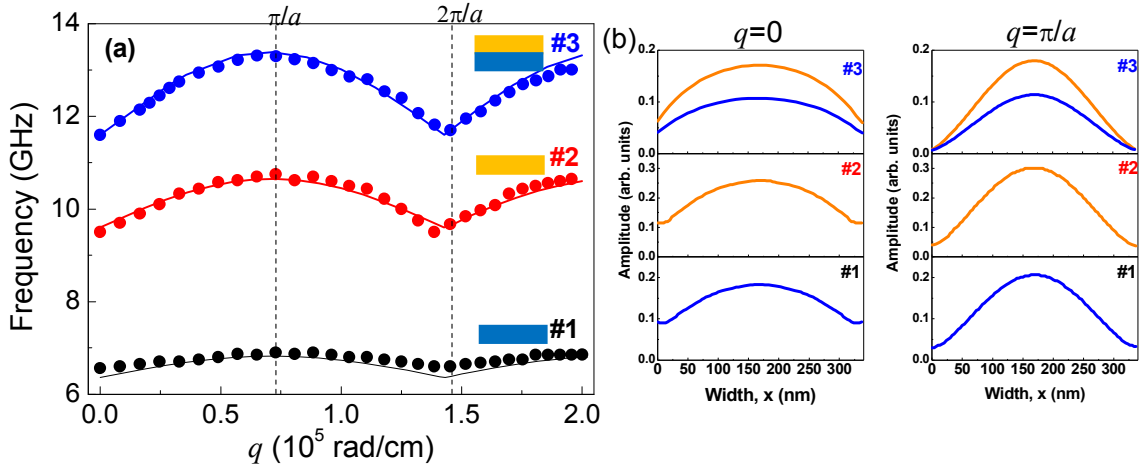

**Fig. 3** (Color online) (a) Measured (points) and calculated (lines) frequency dispersion of the fundamental collective SW mode for the different NWs with rectangular cross-section (sample 1# to #3). The external magnetic field, $H = 500$ Oe, is applied along the length of NWs, while the wave vector is in the direction perpendicular to the NWs length. The vertical dashed lines mark the edge of the first ($\pi/a)$ and second ($2\pi/a)$ Brillouin zones. (b) Calculated profiles of the dynamic magnetization through the NW width for the lowest collective mode, blue (orange) curves are for Py (Fe) layers.

For the bi-layer NWs (Sample #3) the precession is in-phase in the two layers, with a larger amplitude in the Fe layer than in the Py one. To obtain this result we assumed strong ferromagnetic interlayer exchange interaction ($A_{12}>0$) between the layers. This interaction pushes the frequency of the optic mode of the bi-layer structure beyond the highest frequency detected in the experiment (24 GHz). This is different from what previously observed in isolated (non-interacting) Py/Cu/Py NWs[23] where the dipolar coupling through the Cu spacer leads to the appearance of stationary modes of both acoustic (in-phase) and optic (anti-phase) types at relatively low frequencies. An interesting effect that is seen in the spatial profiles of the modes (see Panel b of Fig. 3) is the pronounced amplitude suppression for the NWs with the rectangular cross-section at the NW edges. This effect can be attributed to the effective dipole pinning of magnetization at the edges[35] and to the presence of the collective dynamic field of the array, as discussed in Ref. 23.

*b) Dispersion relation and spatial profiles for the L-shaped NWs array*

Let now consider the frequency dispersion and the spatial profiles for the SW modes in "L-shaped" cross-section NWs. The comparison between the experimental and calculated frequency dispersion for Samples #4 and #5, i.e. for Fe thickness of 10 and 20 nm, is shown in Fig. 4 (a) and (b), respectively, together with the calculated profiles of the most dispersive modes (Fig 4 (c)-(j). One sees that several peaks have been detected by BLS, whose frequencies are downshifted on increasing the thickness of the Fe NWs. Interestingly, we have found that the lowest-frequency mode, oscillates in the frequency range between 8 and 9 GHz (at higher (smaller) frequency than the lowest frequency mode of the single layer Py (Fe) NWs, see Fig. 4). In addition, there is a second mode, at higher frequency which exhibits a sizeable periodic frequency oscillation as a function of the Bloch wave number $q$. The oscillation amplitude of this latter dispersive mode grows when the Fe thickness passes from 10 nm (0.30 GHz) to 20 nm (0.45 GHz). The presence of the two dispersive modes, marked by the shaded regions in Fig. 4, suggests a strong dynamic dipole coupling of individual NWs. This is different from the case of the non-interacting stripes where, as a consequence of lateral confinement, quantized spin waves have been observed. These quantized modes have a stationary character and are dispersionless, i.e. their frequency does not change over the whole range of wave vector investigated.[36,37]

To understand the origin of the dispersive character of the above mentioned modes, it is instructive to look at the spatial profiles of the modes, that are shown in Fig. 4 for $q=0$, Panels (c-f)) and $q=\pi/a$, Panels (g-j). The dynamic magnetization distributions have been averaged over the thicknesses of the respective layers to create the two-dimensional plots. Our first observation is that for the cross-section areas $0<x<170$nm the magnetization precession for both layers is in-phase. Hence in all cases displayed in Fig. 4 (c-j) we deal with the acoustic oscillation of the bi-layer structure Secondly, one sees that the modal profiles for the two Fe layer thicknesses are qualitatively the same, the only difference being the number of anti-nodes of the standing SW across the area of the Py layer not capped with Fe. For the Py(10nm)/Fe(10nm) structure (c,g) the standing SW pattern for this area is close to 1.5 periods of a cosine function while for the Py(10nm)/Fe(20nm) it is close to 2 periods of that function (e,i).

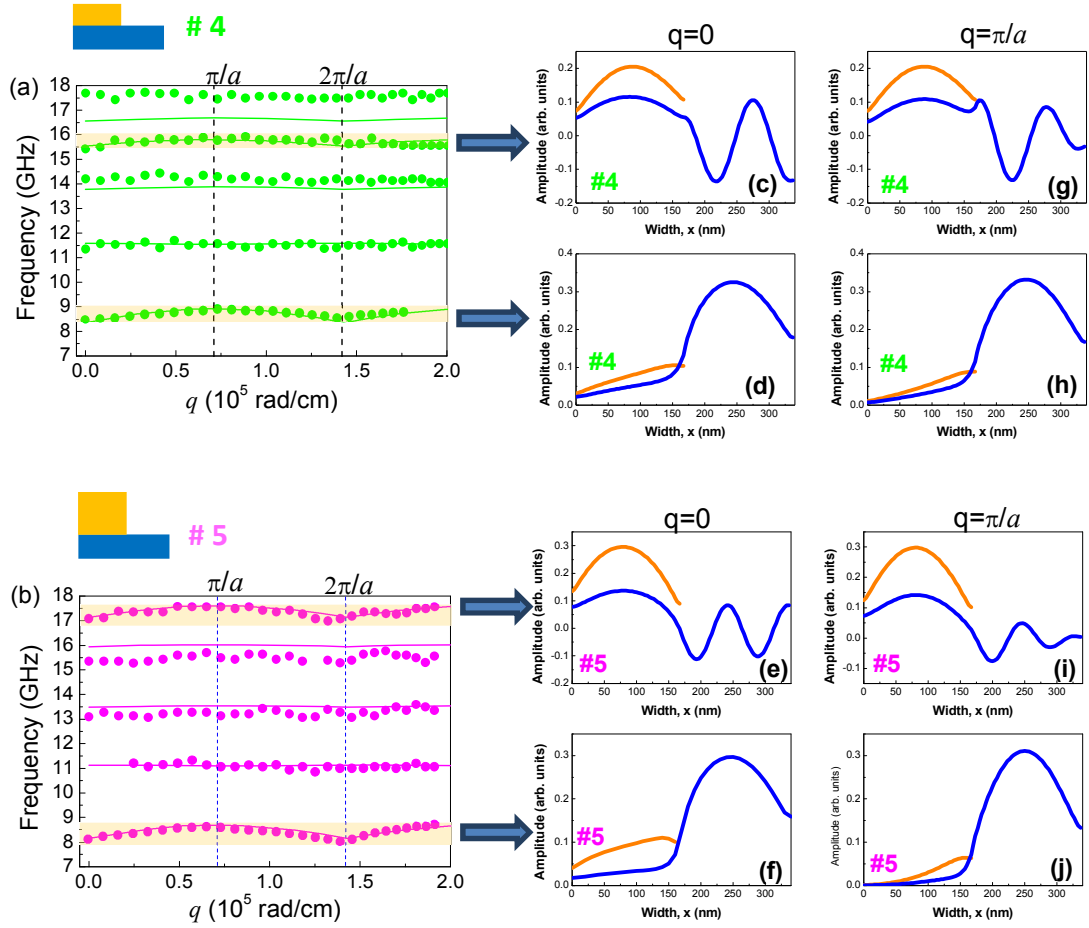

**Fig. 4** (Color online) Measured (points) and calculated (lines) frequency dispersion for L-shaped NWs with Fe thickness of (a) 10 nm (Sample #4) and (b) 20 nm (sample #5). The magnetic field $H$ = 500 Oe is applied along the length of NWs (easy magnetization direction). The vertical dashed lines mark the boundaries of BZs ($q=\pi/a$; $q=2\pi/a$). The amplitude of the magnonic band of the two dispersive modes is represented by shaded regions. (c)-(j) The calculated profiles of the in-plane component of dynamic magnetization for the dispersive modes for the same samples. Blue (orange) curve is for Py (Fe) layer.

A simple way to understand the formation of the modal profiles for the higher-order dispersive mode (Fig. 4(c,e,g,i)) is to use some similarity of this geometry to the case of bi-component ferromagnetic NWs alternated into an array.[38] In this respect, one may represent the L-shape NW as two effective 170 nm-wide NWs. The area 0<$x$<170 nm in the cross-section plane is an effective bi-layer NW consisting of a 10nm-thick Py layer overlaid with either 10 or 20nm-thick Fe layer. We will refer to it as the THICK portion of the NW. The area 170< $x$< 340nm represents an effective 10nm-thick single-layer Py NW, that we will refer to as the THIN portion. The THICK and THIN portions are in a lateral exchange contact through a "virtual interface" placed at $x$=170nm, running across the Py-layer of the actual L-shaped NW. They are also coupled by their dipole fields. One can easily see from the right panels of Fig. 4 that in both samples #4 and #5 the spatial profiles of the THICK portion are qualitatively the same in the two layers, with the main difference is that in sample #5 the frequency of all the modes goes up due to the increased thickness and saturation magnetization of the THICK portion.

Remarkably, one can notice that the only two dispersive modes, whose calculated profiles are reported in Fig. 4, correspond to the fundamental mode of either the THIN or the THICK portions of the NW. In the latter case the frequency is considerably larger, due to the significantly larger thickness and mean saturation magnetization of the THIN portion. This mismatch of the fundamental mode frequencies between the two sides also explains why for the low-frequency fundamental mode of the THICK portion (Fig. 4 d, f, h and j) there is no counterpart to couple in the THIN part. Consequently, the magnetization precession in the latter region is not resonant but represents a forced oscillation (decaying-exponent-like) driven by the exchange coupling through the virtual interface and by the long-range dipole field of the THIN part. On the contrary, the fundamental mode of the THICK portion couples to a much higher individual mode of the THIN one, as seen in the panels c, g, e and i.

Let us now comment on the difference in the profiles for $q$=0 and $q$= $\pi/a$. The most relevant difference, especially in Sample #5, is the different amplitude of the forced oscillation in the Py layer of the THICK region for the low-frequency fundamental mode (compare panels d with h and f with j). The reason for this behavior is that for $q$=0 all the NW on the array resonate in phase, so that the strong dipole field of the fundamental mode generated in both the THICK and the THIN parts penetrates the neighboring NWs (440 < $x$ < 610nm and −440 nm< $x$ < −100 nm). Since the dipole field generated by the neighbors is in-phase with the coupling through the virtual interface $x$=170 nm, it helps to drive the forced magnetization precession in the Py layer of the THICK part. It also enhances the magnetization precession in Py near the edge of the THIN part. As a result, the precession amplitude at both edges is increased. On the contrary, for the lowest frequency mode and $q$=$\pi/a$, the dipole field of the nearest neighbor is in anti-phase with the coupling through the virtual interface. Accordingly, it tends to suppress the forced magnetization oscillation at the edges of the Pylayer of the THICK region. As a result, the profiles of the fundamental mode across the width of the Py layer of NWs look (slightly) more uniform for $q$=0 (Panels d and f) than for $q$= $\pi/a$ (Panels

(h) and (j)). The same argument of the phase of the coupling field can be used in order to explain the differences in the profiles (c),(e) and (g),(i).

*c) Field dependence of the width and the center frequency of the magnonic band*

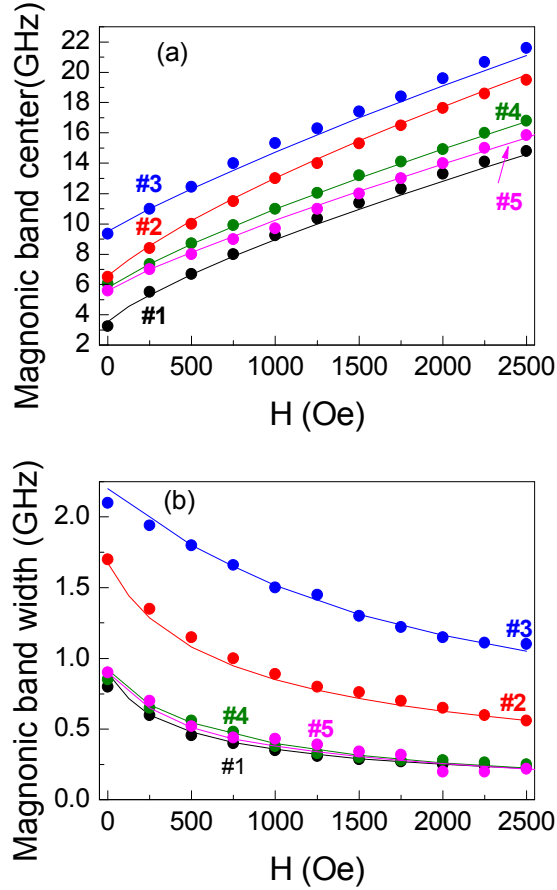

**Fig. 5** (Color online) Comparison between the measured (points) and the calculated (curves) magnonic band center (a) and width (b) of the lowest frequency mode as a function of the magnetic field strength for all the investigated samples (#1-5).

The field dependence of both the width and the center frequency of the magnonic band (amplitude of frequency oscillation) for the lowest frequency mode in all the investigated samples is shown in Fig. 5. First of all we notice that there is a monotonic increase (decrease) of the band center frequency (width) with the intensity of the applied magnetic field. For the three samples with rectangular cross section it is clearly seen that both these quantities plotted in Fig 5. (a) and (b) increase with increasing either the layer magnetization (i.e. passing from Sample #1 to #2) or the film thickness (i.e. passing from Sample #2 to #3). This is because the band center frequency is larger for larger saturation magnetization and thickness, while the band width depends on the strength of the inter-NW dipole coupling (that also increases with magnetization and thickness).

Moreover, one can see that the behavior of the data relative to the L-shaped Samples #4 and #5 is rather similar to that of Sample #1, consisting of a single Py layer. This is because, as discussed in the previous paragraph, the lowest frequency mode in L-shaped samples is strongly localized in the THIN Py side (Panels (d) and (f) of Fig.4). Such mode localization, that would cause a less efficiency coupling with neighbor NWs if compared to Sample #1, is partly compensated in Samples #4 and #5 by the effect of larger average thicknesses and saturation magnetizations. As a result, on can see that the bandwidth for Samples #4 and #5 is practically the same as for #1. In any case, it is remarkable that the calculations accord very well with experimental data, showing that the theoretical model also captures the magnetic field dependent features of the BLS data.

**Conclusions**

In conclusion, the Brillouin light scattering technique has been exploited to measure the dispersion of collective spin waves in arrays of Permalloy/Fe nanowires with either rectangular or L-shaped cross-section, fabricated by a method which combines deep ultraviolet lithography and a self-aligned shadow deposition. The measurements relative to the frequency dispersion of the spin waves have been satisfactorily reproduced by a theory based on the two-dimensional Green's function description of the dynamic dipole field of the oscillating magnetization. This theory also allows us to calculate the spatial profiles of the modes in the two layers thus helping us to comprehend their frequency behavior. For the rectangular nanowires, the fundamental mode lying at the lowest frequency, characterized by an in-phase precession of the magnetization in the two layers and maximum amplitude in the center of the nanowire, exhibits the largest frequency oscillation amplitude. Its frequency in the bi-layer Py/Fe nanowires is up-shifted with respect to that measured for single layer Py and Fe nanowires due to the increased thickness. For the case of the nanowires with L-shaped cross-section, two dispersive modes with sizeable magnonic band have been observed. These are interpreted as the fundamental modes of either the THICK or the THIN portion of each nanowire. The effect of the Fe nanowires width and thickness on magnonic band amplitude and frequency position has also been carried out. We believe that this work can stimulate the design, tailoring, and characterization of magnonic crystals where, thanks to the presence of two contrasting ferromagnetic materials and the inter-layer exchange interaction at their interface, new tailored functionalities can be achieved. We also expect that introducing a thin nonmagnetic spacer between the two ferromagnetic layer and thanks to the different coercivity of the materials, it would be possible to study the spin wave propagation in the case of parallel and anti-parallel alignment of the magnetization in the two layers, thus permitting re-programmable dynamic response of the system.

ACKNOWLEDGMENTS

This work was supported by the MIUR under PRIN Project No. 2010ECA8P3 "DyNanoMag", the Australian Research Council, the University of Western Australia (UWA) and UWA's Faculty of Science and the National Research Foundation, Prime Minister's Office, Singapore under its Competitive Research Programme (CRP Award No. NRF-CRP 10-2012-03).